\documentclass[twocolumn,aps,prl showpacs,floatfix,superscriptaddress]{revtex4-1}
\usepackage{amssymb,amsmath}
\usepackage{hyperref,graphicx}
\usepackage{dcolumn}
\usepackage{color}
\newcommand{\bs}[1]{{\boldsymbol{#1}}}
\newcommand{\bk}{\bs{k}}

\newcommand{\br}{\bs{r}}

\newcommand{\rmd}{\mathrm{d}}

\begin{document}

\title{Thermalization of matter waves in speckle potentials}

\author{Nicolas Cherroret}
\email[Electronic address: ]{cherroret@lkb.upmc.fr}
\affiliation{Laboratoire Kastler Brossel, UPMC-Sorbonne Universit\'es, CNRS, ENS-PSL Research University, Coll\`{e}ge de France, 4 Place Jussieu, 75005 Paris, France}
\author{Tomasz Karpiuk}
\affiliation{Wydzia{\l} Fizyki, Uniwersytet w Bia{\l}ymstoku, ul. Cio{\l}kowskiego 1L, 15-245 Bia{\l}ystok, Poland}
\author{Beno\^it Gr\'{e}maud}
\affiliation{MajuLab, CNRS-UNS-NUS-NTU International Joint Research Unit, UMI 3654, Singapore}
\affiliation{Centre for Quantum Technologies, National University of Singapore, 3 Science Drive 2, Singapore 117543, Singapore}
\affiliation{Department of Physics, National University of Singapore, 2 Science Drive 3, Singapore 117542, Singapore}
\affiliation{Laboratoire Kastler Brossel, UPMC-Sorbonne Universit\'es, CNRS, ENS-PSL Research University, Coll\`{e}ge de France, 4 Place Jussieu, 75005 Paris, France}
\author{Christian Miniatura}
\affiliation{MajuLab, CNRS-UNS-NUS-NTU International Joint Research Unit, UMI 3654, Singapore}
\affiliation{Centre for Quantum Technologies, National University of Singapore, 3 Science Drive 2, Singapore 117543, Singapore}
\affiliation{Department of Physics, National University of Singapore, 2 Science Drive 3, Singapore 117542, Singapore}
\affiliation{INLN, Universit\'{e} de Nice-Sophia Antipolis, CNRS; 1361 route des Lucioles, 06560 Valbonne, France}


\begin{abstract}
We show that the momentum distribution of a nonlinear matter wave suddenly released with a finite velocity in a speckle potential converges, after an out-of-equilibrium evolution, to a universal Rayleigh-Jeans thermal distribution. By exploring the complete phase diagram of the equilibrated wave, we discover that for low but nonzero values of the disorder strength, a large-scale structure --a condensate-- appears in the equilibrium distribution. 
\end{abstract}

\maketitle

Because of their high-degree of control and tunability, quantum gases have become versatile model systems to study effects originating from many different fields of physics such as condensed matter, quantum information, quantum hydrodynamics and even high-energy physics \cite{Lewenstein2007, Bloch2008, Dalibard2011, Thomas2012, Steinhauer2014, Barenghi2014}. The physics of disordered systems does not escape the trend \cite{Clement2006, Shapiro12} with the observation of coherent backscattering and Anderson localization with non-interacting matter waves \cite{Josse12, Billy08, Lemarie09, Jendrzejewski12, Kondov11, Semeghini14}. When interactions are additionally present, disordered gases offer even richer phenomena. For instance, the phase diagram of interacting disordered Bose gases (``dirty boson'' problem) at zero and finite temperatures has recently stirred considerable theoretical and experimental interest \cite{Giamarchi88, Fisher89, Falco09, Aleiner10, Carleo13, Pasienski10, Deissler10, Derrico14}. 

Another important, yet poorly understood problem, is the long-time limit of the out-of equilibrium dynamics of quantum gases released in a disordered potential. 
For spatially-narrow initial states, the atomic cloud spreads out and, in the absence of interactions and under suitable conditions, eventually freezes due to Anderson localization  \cite{Anderson58, Abrahams79}. However, if the matter wave is --even weakly-- interacting, theoretical and experimental evidence suggest that the cloud keeps expanding indefinitely so that no stationary state is ever reached \cite{Kopidakis08, Pikovsky08, Cherroret14}. When atoms are prepared in a plane-wave state at some finite velocity, the interesting dynamics takes place in momentum space \cite{Josse12, Cherroret12, Karpiuk12, Ghosh14, Micklitz14} and the nature of the system at very long times, resulting from the complicated interplay between atomic collisions and scattering off disorder, is largely unknown. This is the question we address in the present Letter. 

Intuitively, the dynamics of weakly-interacting matter waves in momentum space is expected to yield a thermalization process due to atomic collisions. In the homogeneous case, this process has been extensively studied for nonlinear waves obeying the nonlinear Schr\"odinger (NLS) equation, including dilute Bose gases in the mean-field regime, within the framework of weak turbulence theory \cite{Nazarenko11}: the system generically equilibrates to a thermal Rayleigh-Jeans distribution maximizing entropy \cite{Connaughton05, Sun12}. For inhomogeneous systems like disordered systems, this thermalization process needs further analysis since the density of states (DoS) is dramatically altered, in particular at low energies. We have found that, if the equilibrium momentum distribution achieved by a weakly-interacting matter wave evolving in a two-dimensional (2D) speckle potential is still a thermal Rayleigh-Jeans distribution, it also exhibits, for specific values of the disorder strength and of the initial velocity, a large-scale coherent structure. This ``condensate'' coexists with the background of thermalized atoms and disappears at vanishing disorder.

\begin{figure}[h]
\includegraphics[width=0.91\linewidth]{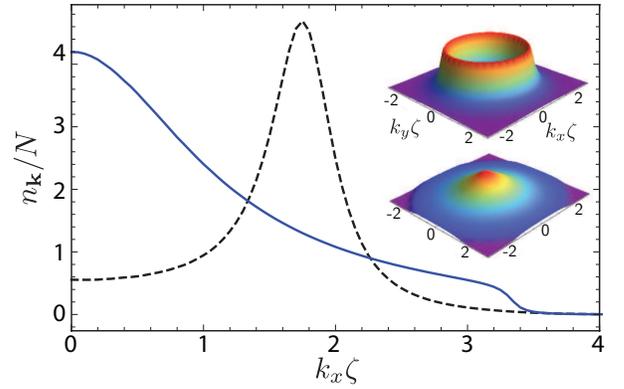}
\caption{Cut at $k_y=0$ of the equilibrium momentum distribution $n_\bk/N$, obtained numerically for $V_0=0.75\epsilon_\zeta$, $E_0=1.5\epsilon_\zeta$ and number density $n=2/\zeta^2$. Dashed curve: $g=0$, see Eq. (\ref{nk_zerog}). Solid curve:  $g\ne0$, see Eq. (\ref{nk_g}). Insets: same distributions shown in the plane $(k_x,k_y)$ for $g=0$ (up) and $g\ne 0$ (down). 
} 
\label{nk_num}
\end{figure}

\emph{Quench scenario --} We consider a gas of $N$ weakly-interacting bosons, initially prepared in the plane-wave state $|\bk_0\rangle$, and suddenly released at $t=0$ in a 2D random potential $V(\br)$ that we chose to be a blue-detuned speckle potential \cite{Clement2006}. Such a random potential has equal mean and root-mean square amplitudes $V_0$ and we use its correlation length $\zeta$ and its correlation energy $\epsilon_\zeta=\hbar^2/(m\zeta^2)$ as units of length and energy.
After the quench, the cloud expands according to the NLS equation, $i\hbar \partial_t\Psi=-\hbar^2/(2m)\boldsymbol{\nabla}^2\Psi+V(\br)\Psi+g|\Psi|^2\Psi$ with $|\Psi|^2$ normalized to the total number of atoms $N$. 
Finally, after a time $t$, the speckle potential is switched off and the disorder-averaged momentum distribution $n_\bk(t)=\overline{|\Psi(\bk,t)|^2}$ of the cloud is recorded. We are interested in the long-time asymptotics $n_\bk(t\to\infty)\equiv n_\bk$. In the absence of interactions ($g=0$), this limit is typically established after the momentum distribution is isotropized by scattering, i. e. after a few transport times $\tau$, and reads \cite{Cherroret12, Karpiuk12, Ghosh14, Micklitz14}:
\begin{equation}
\label{nk_zerog}
\frac{n_\bk}{N}=\int d\epsilon \frac{A_\epsilon(\bk)A_\epsilon(\bk_0)}{\nu(\epsilon)}\ \ \  (g=0).
\end{equation}
Here $A_\epsilon(\bk)$ is the disorder-averaged spectral function of the speckle potential and $\nu(\epsilon) = \int\rmd^2\bk/(2\pi)^2 A_\epsilon(\bk)$ is the disorder-averaged DoS per unit volume, see Supplementary Material. Since $\int\rmd\epsilon A_\epsilon(\bk)=1$, we see that $A_\epsilon(\bk_0)$ is the probability density that an atom with momentum $\bk_0$ has energy $\epsilon$ after the quench. Scattering from the speckle being elastic, this energy distribution remains constant during the evolution and, finally, when the potential is switched off, atoms with energy $\epsilon$ acquire a momentum $\bk$ with probability density $A_\epsilon(\bk)/\nu(\epsilon)$. Note that Eq. (\ref{nk_zerog}) is in fact only valid away from the directions $\pm\textbf{k}_0$, where coherent back and forward scattering interference peaks are expected \cite{Josse12, Cherroret12, Karpiuk12, Micklitz14,Ghosh14}. When $g\ne 0$, these peaks turn out to disappear over a time scale much shorter than the thermalization process discussed in the paper. We will thus not consider them in the following, leaving their study in the presence of interactions for future work.
For an isotropic speckle, $A_\epsilon(\bk)$ only depends on $|\bk|=k$ and $n_{\bk}$ is isotropic as well. 
The momentum distribution (\ref{nk_zerog}) is shown in Fig. \ref{nk_num} for $V_0=0.75\epsilon_\zeta$ and $E_0=\hbar^2k_0^2/(2m)=1.5\epsilon_\zeta$. 
It has a characteristic ring shape reflecting the energy profile of the spectral function \cite{Cherroret12}.

\emph{Thermalization --} When $g\ne0$, atoms are scattered \emph{both} from the fluctuations of $V(\br)$ and of the nonlinear (random) potential $g|\Psi(\br,t)|^2$. The latter process redistributes energies over a collision time scale denoted by $\tau_\text{coll}$. In what follows, we assume $\tau_\text{coll}\gg\tau$, which is fulfilled for small enough $g$. Physically, this condition means that scattering events on the random potential occur more frequently than atomic collisions. Consequently, the disorder isotropizes atomic momenta before the nonlinearity starts to play a role. Thus, when $g\ne0$ the momentum distribution at $t\gg\tau$ is still isotropic, but it keeps evolving:
\begin{equation}
\label{nk_tg}
\frac{n_\bk(t\gg\tau)}{N}=\int d\epsilon A_\epsilon(\bk)\dfrac{f(\epsilon,t)}{n},
\end{equation}
which generalizes Eq.~(\ref{nk_zerog}). The energy distribution $f(\epsilon,t)$ of the matter wave is normalized according to $n=N/\Omega=\int \rmd\epsilon\,\nu(\epsilon)f(\epsilon,t)$, where $\Omega$ is the volume of the system. The change in time of $f(\epsilon,t)$, due to atomic collisions, is controlled by the kinetic equation \cite{Schwiete10, Cherroret11, Schwiete13}
\begin{eqnarray}
\label{coll}
\frac{\partial f(\epsilon,t)}{\partial t}&=&
\int \rmd\hat{{\bf u}}
\prod_{i=2}^4
\rmd\hat{{\bf u}}_i
\rmd\epsilon_i
W_{\epsilon,\epsilon_2,\epsilon_3,\epsilon_4}
\left[(f_{\epsilon}f_{\epsilon_3}f_{\epsilon_4}\right.
\nonumber\\
&&\left.+
f_{\epsilon_2}f_{\epsilon_3}f_{\epsilon_4}
-f_{\epsilon}f_{\epsilon_2}f_{\epsilon_3}
-f_{\epsilon}f_{\epsilon_2}f_{\epsilon_4}\right],
\end{eqnarray}
where we have used the shorthand notation $f_{\epsilon_i}\equiv f(\epsilon_i,t)$ and where $\bk_{\epsilon_i}\equiv\sqrt{2m\epsilon_i}\,\hat{{\bf u}}_i/\hbar$  ($\hat{{\bf u}}_i$ is a unit vector). The collision kernel is $W_{\epsilon,\epsilon_2,\epsilon_3,\epsilon_4}=4\pi\times(2\pi g/\hbar)^2\nu(\epsilon_2)\nu(\epsilon_3)\nu(\epsilon_4)\delta(\bk_\epsilon+\bk_{\epsilon_2}-\bk_{\epsilon_3}-\bk_{\epsilon_4})\delta(\epsilon+\epsilon_2-\epsilon_3-\epsilon_4)$, where the two delta functions stem from momentum and energy conservation during a collision process. Eq. (\ref{coll}) describes the effect of atomic collisions in a ``disordered background'' and reduces to the usual kinetic equation describing short-range interactions between bosons \cite{Griffin09} 
in the homogeneous case where $\nu(\epsilon)=m/(2\pi\hbar^2)$. 
In the absence of interactions, $\partial f(\epsilon,t)/\partial t=0$, whence $f(\epsilon,t)= f(\epsilon,t=0^+)=nA_\epsilon(\bk_0)/\nu(\epsilon)$ and one recovers Eq.~(\ref{nk_zerog}). Furthermore, Eq.~(\ref{coll}) guarantees energy conservation at all times. Expressing it immediately after the quench, and assuming the interaction energy is negligible, we find $E_\text{tot}=\Omega\int d\epsilon\, \epsilon\nu(\epsilon)f(\epsilon,t) = N(V_0+E_0)$. 

The equilibrium energy distribution $f^\text{eq}_\epsilon$ is obtained by canceling the collision kernel and thus solving for $f_{\epsilon}f_{\epsilon_3}f_{\epsilon_4}+f_{\epsilon_2}f_{\epsilon_3}f_{\epsilon_4}-f_{\epsilon}f_{\epsilon_3}f_{\epsilon_4}-f_{\epsilon}f_{\epsilon_2}f_{\epsilon_4}=0$ with $\epsilon=\epsilon_3+\epsilon_4-\epsilon_2$. One finds the Rayleigh-Jeans distribution, $f^\text{eq}_\epsilon=T/(\epsilon-\mu)$. The equilibrium momentum distribution thus reads:
\begin{equation}
\label{nk_g}
\dfrac{n_\bk}{N}=\int\rmd\epsilon\dfrac{A_\epsilon(\bk)}{n}\dfrac{T}{\epsilon-\mu} \ \ \  (g\ne 0),
\end{equation}
which replaces Eq. (\ref{nk_zerog}). The blue-detuned speckle potential being bounded from below by zero \cite{Falco10}, energies start from $0$ in the integral. Since $f_\epsilon^\text{eq}>0$, we must have $T>0$ and $\mu<\text{min}(\epsilon)=0$. The parameters $T$ and $\mu$ can thus naturally be interpreted as the equilibrium \emph{temperature} and \emph{chemical potential} of the cloud after its out-of-equilibrium evolution. They are obtained from atom number and total energy conservation:

\begin{eqnarray}
\label{Tmu_def}
&&n=T\int_0^{E_\text{max}}\!\!\rmd\epsilon\,\dfrac{\nu(\epsilon)}{\epsilon-\mu}, \ \text{and}\\
\label{Tmu_def2} 
&&n(V_0+E_0)=T\int_0^{E_\text{max}}\!\!\rmd\epsilon\,\epsilon\dfrac{\nu(\epsilon)}{\epsilon-\mu}.
\end{eqnarray}
As $\nu(\epsilon\rightarrow\infty)=m/(2\pi\hbar^2)$, both integrals have an ultraviolet (UV) divergence and an upper cutoff $E_\text{max}$ has been introduced for regularization. 
This divergence stems from the mean-field description of interactions, encapsulated by the NLS equation, which provides a classical nonlinear wave model of the gas only valid at low energies.  Following \cite{Nazarenko11}, we proceed by analogy with a non-interacting Bose gas at equilibrium and use the cutoff $E_\text{max}=T+\mu$, which physically describes the cross-over region separating the classical-field and particle-like (Boltzmann) \cite{Geiger12} descriptions of the gas. 

Once Eqs. (\ref{Tmu_def}) and (\ref{Tmu_def2}) have been numerically solved for $T$ and $\mu$, for given values of $E_0$, $V_0$ and $n$, the thermalized momentum distribution follows from Eq. (\ref{nk_g}). The result is shown in Fig.~\ref{nk_num} 
for $E_0=1.5\epsilon_\zeta$, $V_0=0.75\epsilon_\zeta$ and $n=2/\zeta^2$. Unlike for $g=0$, the distribution is now centered at $k=0$ and has a smooth shape.  (except in the far tails, whose details depend on the UV cutoff). 
\begin{figure}[h]
\includegraphics[width=0.99\linewidth]{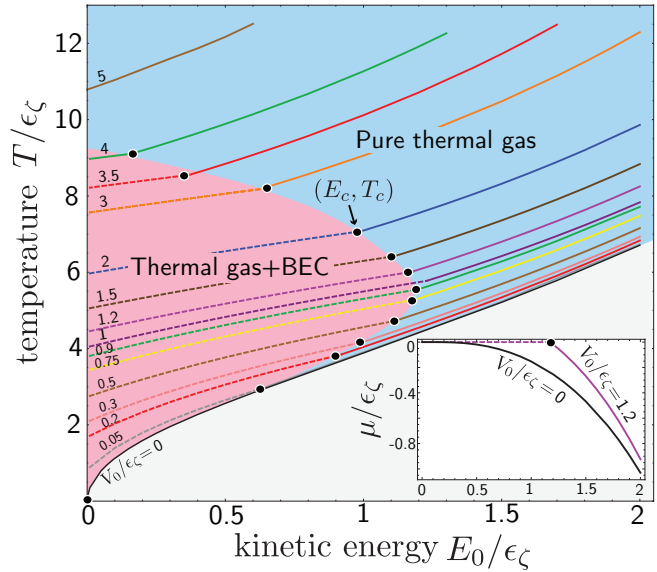}
\caption{Main panel: equilibrium phase diagram $(E_0,T)$ obtained from Eqs. (\ref{Tmu_def}) and (\ref{Tmu_def2}) for various values of $V_0$ at number density $n=2/\zeta^2$. We find a critical curve (black dots) separating a thermal region (blue region, $\mu <0$) from a region where condensation occurs (red region, $\mu = 0$). Solid curves: temperature $T$ as a function of $E_0$ in the thermal region. Dashed curves: temperature T as a function of $E_0$ in the region where condensation occurs. 
The value of $V_0$ is indicated on top of each equipotential. The lowest black curve ($V_0=0$) reproduces the scenario of weak turbulence in two dimensions, for which there is no condensation at finite temperature ($T_c=0$). Inset: chemical potential $\mu$ as a function of $E_0$, for $V_0=0$ and $1.2\epsilon_\zeta$. When $V_0\ne0$, $\mu$ vanishes at $E_0=E_c$.
} 
\label{diagram_TE0}
\end{figure}

\emph{Condensation --} We now explore in more details 
the $(T,E_0)$-phase diagram of the system 
for different disorder strengths $V_0$ at fixed number density. Fig. \ref{diagram_TE0} shows our results obtained by solving numerically Eqs.~(\ref{Tmu_def}) and (\ref{Tmu_def2}) for several values of $V_0$ and $E_0$ at $n=2/\zeta^2$.
For values of $V_0$ and $E_0$ in the red region, no solution 
fulfilling the constraints $T>0$ and $\mu<0$ can be found. The boundary of this red region (black dots) defines a critical curve $E_c(V_0,n)$ where the chemical potential $\mu$ vanishes (the inset shows the vanishing of $\mu$ for $V_0=1.2\epsilon_\zeta$). To each value of $E_c$ corresponds in turn a certain critical temperature $T_c(V_0,n)$. Physically, this critical line signals that at fixed ($V_0,n$), there is a saturation of the population of excited energy levels, which can no longer accommodate particles if $E_0$ is decreased below $E_c$ (and thus when $T$ becomes smaller than $T_c$). This phenomenon is characteristic of the appearance of a Bose-Einstein condensate (BEC) in the ground state of the speckle potential \cite{Castin00}. 
For $T\leq T_c$, the gas thus consists of a thermal part coexisting with $N_\text{BEC}$ condensed atoms. In this phase, the total energy of the system is still given by Eq. (\ref{Tmu_def2}), but with $\mu=0$ as in the usual Bose condensation. Solving this equation then gives access to the temperature for $E_0\leq E_c$ and is shown in Fig. \ref{diagram_TE0} (dashed curves).
Let us briefly discuss the shape of this low-temperature phase. First, condensation in the speckle potential exists only at low enough values of $E_0$ and $V_0$. The reason is that at large $E_0$ or $V_0$, too much energy is injected in the system at $t=0$, and the final equilibrium temperature is correspondingly too large for condensation to be possible. Second, it should be noted that the presence of the random potential is \emph{crucial} for the emergence of a BEC. Indeed, in the limit of vanishing disorder, obtained by setting $V_0=0$ and $\nu(\epsilon)=m/(2\pi\hbar^2)$ in Eqs. (\ref{Tmu_def}) and (\ref{Tmu_def2}), $\mu$ vanishes only for $E_0=0$ (see the inset of Fig. \ref{diagram_TE0}), and $T_c\to 0$: no BEC ever appears. In fact, this limit coincides with the scenario of weak turbulence of nonlinear waves \cite{Nazarenko11}, for which it is well known that there is no BEC formation in 2D homogeneous systems except at T=0 \cite{Connaughton05}. The reason for this difference lies in the behavior of the DoS, which vanishes at low energies in the presence of disorder, while it is always constant in the absence of disorder.

\begin{figure}[h]
\includegraphics[width=0.91\linewidth]{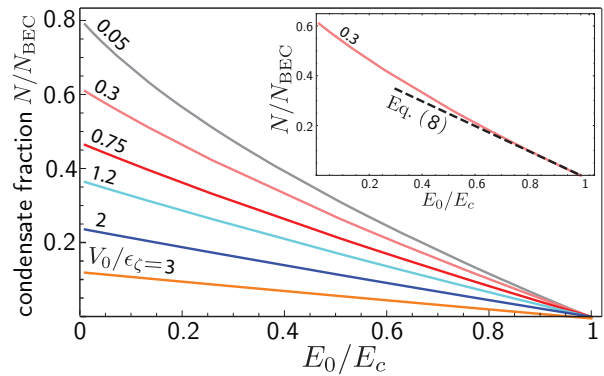}
\caption{Condensate fraction as a function of $E_0$ (in units of the critical energy $E_c(V_0,n)$ where condensation occurs) for several values of $V_0$. The total number density is $n=2/\zeta^2$.
Inset: comparison between the condensate fraction 
at $V_0=0.3\epsilon_\zeta$ and the theoretical prediction (\ref{F_near_ec}) valid near $E_0=E_c$ (dashed line).} 
\label{frac}
\end{figure}

To find the fraction of condensed atoms when $T\leq T_c$, we must update Eq.~(\ref{nk_g}) 
by adding the BEC component:

\begin{equation}
\label{mom_tot}
\dfrac{n_\bk}{N}=\dfrac{n_\text{BEC}(\bk)}{N}+\int_0^T\rmd\epsilon \dfrac{A_\epsilon(\bk)}{n}\dfrac{T}{\epsilon}\ \ (T\leq T_c),
\end{equation}
where $n_\text{BEC}(\bk)$ is the condensate momentum distribution with $\int\rmd^2\bk/(2\pi)^2 \, n_\text{BEC}(\bk)=N_\text{BEC}$. Integrating over $\bk$, we get the fraction of condensed atoms $N_\text{BEC}/N=1-(T/n)\int_0^T\rmd\epsilon\,\nu(\epsilon)/\epsilon$. Once $T$ is known from Eq.~(\ref{Tmu_def2}) with $\mu=0$, one can use Eq.~(\ref{mom_tot}) to find $N_\text{BEC}/N$, see Fig.~\ref{frac}.
 
We see that the condensate fraction is always smaller than unity: 
because of the quench at $t=0$, the average energy $V_0$ is suddenly injected to the gas, which eventually leads to a rather high equilibrium temperature. 
By expanding the condensate fraction around $T=T_c$ and eliminating $T$, we obtain the following theoretical prediction valid for $E_0$ close to $E_c(V_0,n)$:
\begin{equation}
\label{F_near_ec}
\frac{N_\text{BEC}}{N} \simeq\dfrac{E_c-E_0}{E_c}\dfrac{1+T_c\nu(E_c)/n}{1+V_0/E_c+T_c^2\nu(E_c)/(n E_c)}. 
\end{equation}
Prediction~(\ref{F_near_ec}) is compared to the exact result at $V_0=0.3\epsilon_\zeta$ and $n=2/\zeta^2$ in the inset of Fig. \ref{frac}.

To evaluate the condensate momentum distribution $n_\text{BEC}(\bk)$ and its typical width $\Delta k$, and further find $n_\bk$ when  $T\leq T_c$, we proceed by analogy with the study of the ground state of a Bose gas at equilibrium in a confining potential \cite{Falco09}. Condensation takes place in the lowest energy state of the system which is found, since the blue-detuned speckle potential is bounded from below, in the largest potential well, of typical size $R\sim 1/\Delta k$. If the system size $\sqrt{\Omega}$ is large enough, the condensed state lies in the Lifshitz tail of  the density of states where $\nu(\epsilon)\sim \exp(-\epsilon_\zeta/\epsilon)$ \cite{Lifshitz1964}, and its energy is typically $\epsilon(R)\sim\hbar^2/(2m R^2)$. The radius $R$ of the condensed state is such that $\Omega \int_0^{\epsilon(R)}\rmd\epsilon\,\nu(\epsilon)=1$. 
This gives $\Delta k\sim1/R\sim\zeta^{-1}\ln^{-1/2}(\sqrt{\Omega}/\zeta)$. 
From the normalization of $n_\text{BEC}(\bk)$, we have $n_\text{BEC}(\bk=0)(\Delta k)^2\sim N_\text{BEC}$, such that $n_\text{BEC}(\bk=0)\sim \zeta^2\ln(\sqrt{\Omega}/\zeta)$. Thus, in the thermodynamic limit $\sqrt{\Omega}\to \infty$, we see that the momentum distribution for $T\leq T_c$ consists of a narrow peak at $\bk=0$, $n_\text{BEC}(\bk)\to N_\text{BEC}\delta(\bk)$, sitting on top of the Rayleigh-Jeans background of thermal atoms.

\emph{Time scale of thermalization --}
We now address the question of the time scale $\tau_\text{coll}$ needed to achieve \emph{thermal} equilibrium.
To this end we consider a small perturbation $f_\epsilon(t)=f_\epsilon^\text{eq}+A(t)\delta(\epsilon)$ 
and look at its exponential relaxation $A(t)=A(0)\exp{(-t/\tau_\text{coll})}$ when substituted in the linearized kinetic equation obtained from ($\ref{coll}$). This procedure, detailed in the Supplementary Material, leads to $\tau_\text{coll}/\tau_\zeta=[\hbar^2/(gm)]^2F(E_0,V_0,n)$, where $\tau_\zeta=m\zeta^2/\hbar$.
The dimensionless function $F$ is smooth as long as one is not too close to the condensation threshold. For $E_0=1.5\epsilon_\zeta$, $V_0=0.2\epsilon_\zeta$ and $n=2/\zeta^2$, the gas is in the thermal phase, see Fig. \ref{diagram_TE0}, and we find $F(E_0,V_0,n)\simeq 0.6$. For a quasi-2D gas of $^{39}$K atoms 
with a typical s-wave scattering length $a=200a_0$ ($a_0=53$pm) \cite{Deissler10} and confined in the transverse direction by a harmonic potential of frequency $\nu_\perp=100$Hz, we find $\tau_\text{coll}\simeq 550\tau_\zeta$. Taking the value $\zeta=0.2\mu$m, achieved in recent experiments \cite{Jendrzejewski12}, we obtain $\tau_\text{coll}\simeq 14$ ms, a value much smaller that the typical duration of experiments (of the order of a few seconds). This makes the observation of the thermalization process described in this Letter accessible to current experimental setups.

\emph{Conclusion --}
We have shown that the momentum distribution of a nonlinear matter wave released with finite velocity in a 2D speckle potential undergoes a thermalization process to a Rayleigh-Jeans distribution. 
At low enough (but nonzero) kinetic energy and disorder strength, a condensation process also takes place at $\bk=0$.  These predictions should be within reach of cold atoms experiments. 
Future studies could include a detailed analysis of the dynamics of thermalization and condensation, and address the effect of residual interactions in the equilibrated state, which for static systems at equilibrium lead to the BKT physics \cite{Carleo13}. Finally, it would be interesting to consider 
the case where the initial state is no longer a plane wave but a wave packet. This would allow to establish a connection between the dynamics in momentum space and in configuration space \cite{Kopidakis08, Pikovsky08, Cherroret14}.


NC and ChM aknowledge invaluable discussions with Dominique Delande and Denis Basko. The Centre for Quantum Technologies is a Research Centre of Excellence funded by
the Ministry of Education and National Research Foundation of Singapore.


\end{document}


\title{Supplementary material: Thermalization of nonlinear matter waves in speckle potentials}

\author{Nicolas Cherroret}
\email[Electronic address: ]{cherroret@lkb.upmc.fr}
\affiliation{Laboratoire Kastler Brossel, UPMC-Sorbonne Universit\'es, CNRS, ENS-PSL Research University, Coll\`{e}ge de France, 4 Place Jussieu, 75005 Paris, France}
\author{Tomasz Karpiuk}
\affiliation{Wydzia{\l} Fizyki, Uniwersytet w Bia{\l}ymstoku, ul. Cio{\l}kowskiego 1L, 15-245 Bia{\l}ystok, Poland}
\author{Beno\^it Gr\'{e}maud}
\affiliation{MajuLab, CNRS-UNS-NUS-NTU International Joint Research Unit, UMI 3654, Singapore}
\affiliation{Centre for Quantum Technologies, National University of Singapore, 3 Science Drive 2, Singapore 117543, Singapore}
\affiliation{Department of Physics, National University of Singapore, 2 Science Drive 3, Singapore 117542, Singapore}
\affiliation{Laboratoire Kastler Brossel, UPMC-Sorbonne Universit\'es, CNRS, ENS-PSL Research University, Coll\`{e}ge de France, 4 Place Jussieu, 75005 Paris, France}
\author{Christian Miniatura}
\affiliation{Merlion MajuLab, CNRS-UNS-NUS-NTU International Joint Research Unit, UMI 3654, Singapore}
\affiliation{Centre for Quantum Technologies, National University of Singapore, 3 Science Drive 2, Singapore 117543, Singapore}
\affiliation{Department of Physics, National University of Singapore, 2 Science Drive 3, Singapore 117542, Singapore}
\affiliation{INLN, Universit\'{e} de Nice-Sophia Antipolis, CNRS; 1361 route des Lucioles, 06560 Valbonne, France}

\maketitle

\appendix
\section{Spectral function and density of states in a 2D speckle potential}

In our numerical simulations, $V(\textbf{r})$ is chosen to be a 2D, blue-detuned speckle potential with mean value $\av{V(\textbf{r})}=V_0$ and two-point correlator $\av{\delta V(\textbf{r})\delta V(\textbf{r}')}=[2V_0J_1(|\textbf{r}-\textbf{r}'|/\zeta)/(|\textbf{r}-\textbf{r}'|/\zeta)]^2$, where $\delta V({\bf r}) = V({\bf r}) -V_0$ and $\zeta$ is the correlation length. This potential is numerically generated by convoluting a circular Gaussian random field with a cutoff function  that simulates the diffusive plate used in experiments \cite{Huntley89, Horak98}. The disorder-averaged spectral function, $A_\epsilon(\bk_0)=\int (dt/\hbar) \langle\bk_0|\overline{\text{exp}(-i\hat{H}t/\hbar)} |\bk_0\rangle\text{exp}(i \epsilon t/\hbar)$, is obtained by propagating the plane wave state $|\bk_0\rangle$ with the Hamiltonian $\hat{H}=\hat{\textbf{p}}^2/(2m)+V$, taking the Fourier transform of the result with respect to time and redoing the same calculation for many disorder configurations (as indicated by the over bar) \cite{footnote}. The time propagation is achieved on a 2D grid of size $L\times L$ with periodic boundary conditions along $x$ and $y$, by using an iterative method based on the expansion of the evolution operator $\exp(-i\hat{H}t/\hbar)$ in combinations of Chebyshev polynomials of the Hamiltonian \cite{Fehske09, Roche97}. 
\begin{figure}[h]
\includegraphics[width=0.9\linewidth]{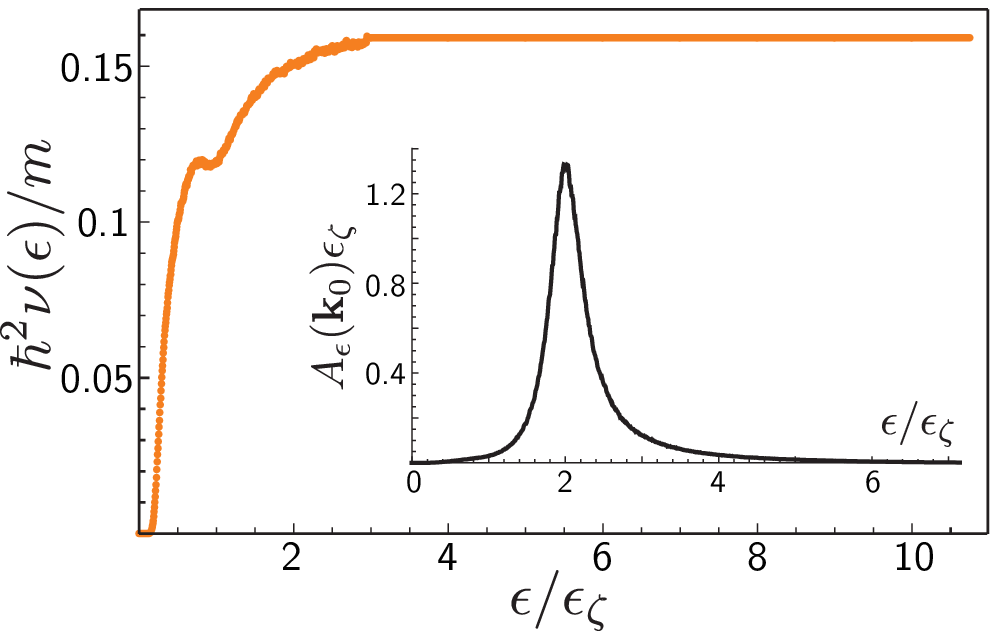}
\caption{
2D density of states $\nu(\epsilon)$ in a speckle potential, for $V_0=0.75\epsilon_\zeta$. Inset: spectral function $A_\epsilon(\bk_0)$, for $V_0=0.75\epsilon_\zeta$ and $|\bk_0|=1.7/\zeta$.
}
\label{dos_V075}
\end{figure}
In the simulations, a cell of surface $(\pi\zeta)^2$ is discretized in typically 8-10 steps along both axis $x$ and $y$ and results are averaged over 8 disorder realizations. The inset of Fig. \ref{dos_V075} displays a typical plot of the disorder-averaged spectral function $A_\epsilon(\bk_0)$ as a function of the energy $\epsilon$, for $V_0=0.75\epsilon_\zeta$ and $|\bk_0|=1.7/\zeta$ 
with $\epsilon_\zeta=\hbar^2/(m\zeta^2)$ the correlation energy of the speckle potential.

Finally, the disorder-averaged density of states per unit volume in the speckle potential, $\nu(\epsilon)$, is obtained by computing spectral functions at different values of $\bk$, and then summing then over $\bk$. Fig. \ref{dos_V075} shows $\nu(\epsilon)$ obtained with this procedure for $V_0=0.75\epsilon_\zeta$. As expected, $\nu(\epsilon)$ is zero for $\epsilon<0$ (the blue-detuned speckle potential is always positive), and $\nu(\epsilon)$ converges to the free-space result $m/(2\pi\hbar^2)$ at large energies.

\section{Estimation of the thermalization time}

From the kinetic equation (3), we can estimate the collision time $\tau_\text{coll}$, which gives a lower bound for the time needed by the system to reach the Rayleigh-Jeans thermal distribution $f_\epsilon^\text{eq}=T/(\epsilon-\mu)$ when $g\ne0$. For this purpose, we consider a small perturbation from the equilibrium distribution at a certain energy $\epsilon_0$:
\begin{equation}
\label{f_ansatz}
f_\epsilon(t)=f_\epsilon^\text{eq}+A(t)\delta(\epsilon-\epsilon_0).
\end{equation}
Substituting Eq. (\ref{f_ansatz}) for all the $f_{\epsilon_i}$ in Eq. (3) and linearizing in $A(t)$, we obtain, after performing integrals over $\hat{\textbf{u}}$, $\hat{\textbf{u}}_2$, $\hat{\textbf{u}}_3$ and $\hat{\textbf{u}}_4$:
\begin{eqnarray}
\label{dA_dt}
&&\dfrac{\rmd A(t)}{\rmd t}\delta(\epsilon-\epsilon_0)=8 g^2\hbar A(t)\int_0^{T+\mu}
\rmd\epsilon_2
\rmd\epsilon_3
\rmd\epsilon_4
\nonumber\\
&&\nu(\epsilon_2)\nu(\epsilon_3)\nu(\epsilon_4)
\delta(\epsilon+\epsilon_2-\epsilon_3-\epsilon_4)
\chi(k_\epsilon,k_{\epsilon_2},k_{\epsilon_3},k_{\epsilon_4})\nonumber\\
&&\times\big\{\delta(\epsilon_4-\epsilon_0)\left[f_{\epsilon_3}(f_{\epsilon}+f_{\epsilon_2})-f_\epsilon f_{\epsilon_2}\right]+\nonumber\\
&&\ \ \,\delta(\epsilon_3-\epsilon_0)\left[f_{\epsilon_4}(f_{\epsilon}+f_{\epsilon_2})-f_\epsilon f_{\epsilon_2}\right]+\nonumber\\
&&\ \ \,\delta(\epsilon-\epsilon_0)\left[f_{\epsilon_3} f_{\epsilon_4}-f_{\epsilon_2}(f_{\epsilon_3}+f_{\epsilon_4})\right]+\nonumber\\
&&\ \ \,\delta(\epsilon_2-\epsilon_0)\left[f_{\epsilon_3} f_{\epsilon_4}-f_{\epsilon}(f_{\epsilon_3}+f_{\epsilon_4})\right]\big\}.
\end{eqnarray}
The function $\chi$ has been explicitly calculated in \cite{Schwiete13}. For clarity we do not reproduce its expression here. After integration of both sides of Eq. (\ref{dA_dt}) over an infinitesimal energy interval centered around $\epsilon_0$, only the third term in the right-hand side survives, which yields
\begin{equation}
A(t)\propto\exp\left[-\dfrac{t}{\tau_\text{coll}(\epsilon_0)}\right],
\end{equation}
with
\begin{eqnarray}
\label{taucoll_general}
\dfrac{1}{\tau_\text{coll}(\epsilon_0)}&=&8 g^2\hbar \int_0^{T+\mu}
\rmd\epsilon_2
\rmd\epsilon_3
\rmd\epsilon_4
\nu(\epsilon_2)\nu(\epsilon_3)\nu(\epsilon_4)\nonumber\\
&&\times\delta(\epsilon_0+\epsilon_2-\epsilon_3-\epsilon_4)
\chi(k_{\epsilon_0},k_{\epsilon_2},k_{\epsilon_3},k_{\epsilon_4})\nonumber\\
&&\times\left[f_{\epsilon_3} f_{\epsilon_4}-f_{\epsilon_2}(f_{\epsilon_3}+f_{\epsilon_4})\right],
\end{eqnarray}
where we have used the same ultraviolet cutoff as in the main article.
Eq. (\ref{taucoll_general}) is the general expression for the collision time as a function of the energy $\epsilon_0$. Since we only aim at giving an estimation of $\tau_\text{coll}$, we focus on the particular energy $\epsilon_0=0$ (where the Rayleigh-Jeans equilibrium energy distribution is maximum). Expliciting $\chi$ \cite{Schwiete13} and performing the integral over $\epsilon_2$, Eq. (\ref{taucoll_general}) simplifies to
\begin{equation}
\label{taucoll_E0}
\dfrac{\tau_\text{coll}(0)}{\tau_\zeta}=\left(\frac{\hbar^2}{gm}\right)^2F(E_0,V_0,n),
\end{equation}
where $\tau_\zeta=m\zeta^2/\hbar$. The inverse of the function $F$ is given by:
\begin{eqnarray}
\label{f}
&&F^{-1}(E_0,V_0,n)=
\int_0^{T+\mu}\rmd x
\int_0^{T+\mu-x}\rmd y
\tilde\nu(x+y)\tilde\nu(x)\tilde\nu(y)\times\nonumber\\
&&
\dfrac{-4\pi\tilde\mu \tilde T^2\Theta(\sqrt{x}+\sqrt{y}-\sqrt{x+y})\Theta(\sqrt{x+y}-|\sqrt{x}-\sqrt{y}|)}{\sqrt{x y}(x-\mu)(y-\mu)(x+y-\mu)},\nonumber\\
&& 
\end{eqnarray}
where $\Theta$ is the Heaviside function and where we have introduced the dimensionless density of states $\tilde\nu(\epsilon)=\hbar^2\nu(\epsilon)/m$, temperature $\tilde T=T/\epsilon_\zeta$ and chemical potential $\tilde\mu=\mu/\epsilon_\zeta$. 

\begin{figure}
\includegraphics[width=0.95\linewidth]{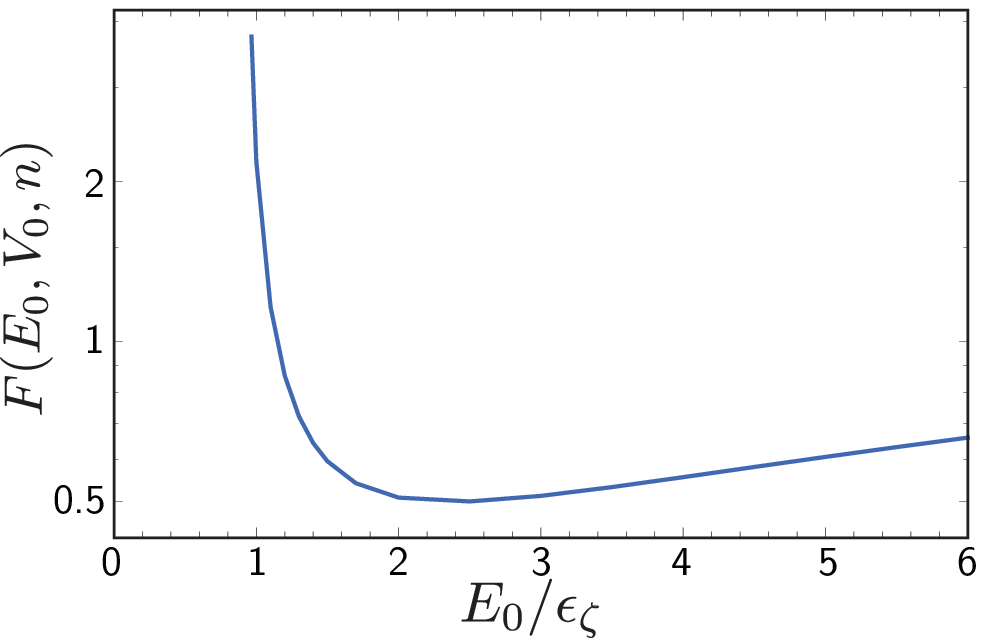}
\caption{Plot of $F(E_0,V_0,n)$ as a function of $E_0$, for $V_0=0.2\epsilon_\zeta$ and $n=2/\zeta^2$.} 
\label{taucoll}
\end{figure}
Fig. \ref{taucoll} displays $F(E_0,V_0,n)$ as a function of $E_0$, for $V_0=0.2\epsilon_\zeta$ and $n=2/\zeta^2$. $F$ increases with $E_0$ at large enough $E_0$, which is consistent with the intuitive picture that it is harder to thermalize when more energy is injected in the system. The sudden increase of $F$ at small $E_0$ signals the approach to the BEC transition, which for $V_0=0.2\epsilon_\zeta$ and $n=2/\zeta^2$ takes place at $E_0=E_c\simeq0.9\epsilon_\zeta$ (see Fig. 2 of the main article). Note that the knowledge of $F$ for $E\leq E_c$ would require a more general kinetic equation describing exchanges between condensed and non-condensed particles, a task which is beyond the scope of the present work. From Fig. \ref{taucoll}, we find $F(E_0,V_0,n)\simeq 0.6$ for $E_0= 1.5\epsilon_\zeta$, which is the value used in the main article. 

For completeness we mention  that $F\propto (n\zeta^2)^{-2}$ at low density, a property that we have checked numerically. Making use of Eq. (\ref{taucoll_E0}), we thus find that the condition $\tau_\text{coll}\gg\tau$, expressing the validity of our kinetic approach, reduces to $[\hbar^2/(gmn\zeta^2)]^2\gg \tau/\tau_\zeta$. The speckle scattering time $\tau$ has been calculated in \cite{Kuhn07} for weak disorder, where it was found that $\tau/\tau_\zeta\sim(\epsilon_\zeta/V_0)^2$ at low energies. With this expression, and overlooking the $V_0$-dependence of $F$, the condition of weak interactions quite naturally reads $gn\ll V_0$.